\documentclass[pra,twocolumn,showpacs,superscriptaddress,floatfix]{revtex4}
\usepackage[dvipdfmx,dvipdfmx]{graphicx}
\usepackage{graphicx}
\usepackage{color}

\usepackage{subfigure}
\usepackage{epsfig}
\usepackage{float}
\usepackage{epstopdf}

\usepackage{amsmath}
\usepackage{amsfonts}
\newcommand{\abs}[1]{\left| #1 \right|}
\newcommand{\ket}[1]{\left | #1 \right \rangle}
\def\k(#1){|#1\rangle}
\newcommand{\bra}[1]{\left \langle #1 \right |}

\newcommand{\tr}{{\rm \, Tr }\, }

\newcommand{\beq}{\begin{equation}}
\newcommand{\eeq}{\end{equation}}
\newcommand{\beqa}{\begin{eqnarray}}
\newcommand{\eeqa}{\end{eqnarray}}
\newcommand{\beqan}{\begin{eqnarray*}}
\newcommand{\eeqan}{\end{eqnarray*}}

\newcommand{\affA}{%
\affiliation{
 Center for Macroscopic Quantum States (bigQ), Department of Physics, Technical University of Denmark, Building 309, 2800 Lyngby, Denmark}
     }

\begin{document}

\title{Tomography of a displacement photon counter for discrimination of single-rail optical qubits 
}


\author{Shuro Izumi}
\affA
\author{Jonas S. Neergaard-Nielsen}%
\affA
\author{Ulrik L. Andersen}%
\affA

\begin{abstract}
We investigate the performance of a detection strategy composed of a displacement operation and a photon counter, which is known as a beneficial tool in optical coherent communications, to the quantum state discrimination of the two superpositions of vacuum and single photon states corresponding to the $\hat\sigma_x$ eigenstates in the single-rail encoding of photonic qubits.
We experimentally characterize the detection strategy in vacuum-single photon two-dimensional space using quantum detector tomography and evaluate the achievable discrimination error probability from the reconstructed measurement operators.
We furthermore derive the minimum error rate obtainable with Gaussian transformations and homodyne detection. 
Our proof of principle experiment shows that the proposed scheme can achieve a discrimination error surpassing homodyne detection.
\end{abstract}

\maketitle

\section{Introduction}\label{Sect:1}
Discrimination of quantum states is an essential task in various quantum information processing (QIP) applications. 
A discrimination error induced as a result of the inherent quantum uncertainty of the interrogated quantum states and/or an inadequate quantum measurement strategy limits the performance of QIP.
Therefore, the discrimination of the quantum states with a minimum error probability is often required to accomplish the application with optimal performance.
Measurement in quantum physics is represented by a positive operator valued measure (POVM) and the POVM representation of the optimal measurement that minimize the discrimination error have been well studied \cite{Helstrom_book76_QDET}.
However, the physical implementations of the optimal measurements are not always trivial and designing the physical detection structure by combining the available resources is another important aspect of the quantum state discrimination.


One of the principal examples that requires a non-trivial quantum measurement is the discrimination of the superposition of the vacuum and the single photon states, $\{\ket{+}=(\ket{0}+\ket{1})/\sqrt{2}, \ket{-}=(\ket{0}-\ket{1})/\sqrt{2}\}$.
For QIP associated with the single-rail optical qubit \cite{BabichevPRL,UlrikRalph,UlrikJonas,Ulanov}, the logical bases $\{\ket{0}_L = \ket{0}, \ket{1}_L = \ket{1}\}$ can be easily discriminated by an ideal photon counter.
On the other hand, discriminating the conjugate basis states $\{\ket{+}, \ket{-}\}$ requires a more complicated procedure even though an error-free discrimination is, in principle, possible owing to the orthogonality of the states.
If a Hadamard gate operation for the single-rail qubit were readily available, such perfect discrimination would be straight-forward as the gate converts $\ket{+}$ into $\ket{0}$ and $\ket{-}$ into $\ket{1}$ \cite{LundRalph,RalphLundWiseman,Laghaout}.
However, a deterministic Hadamard gate requires very strong optical nonlinearities in order to create and annihilate single photons and is therefore infeasible.
It is therefore of interest to explore physically feasible measurement that more directly projects a quantum state onto the $\ket{\pm}$ basis.

In this article, we study the efficiency of using a simple displacement-based photon counter for the discrimination of $\ket{+}$ and $\ket{-}$.  
Such a measurement procedure, which we call a displacement photon counter hereafter, was originally proposed for binary phase shift keying (BPSK) coherent states discrimination.
Here it was shown to provide a means of discriminating coherent states that outperforms the standard homodyne detector strategy in a certain photon number regime \cite{Kennedy73}.
Since its inception, a lot of attention has been devoted to the investigation of the receiver from both a theoretical \cite{Dolinar73,Bondurant93,sasakihirota,TakeokaSasakiLutkenhaus2006_PRL_BinaryProjMmt,Takeoka2008,Takeokavisi,izumi2012,izumi2013} and an experimental point of view \cite{CookMartinGeremia2007_Nature,Wittmann2008_PRL_BPSK,Tsujino2010_OX_OnOff,Tsujino2011_Q_Receiver_BPSK,Muller2012_NJP,Becerra13,Becerra15}.
In contrast, very little work has been devoted to investigating the potential of applying the displacement photon counter to discriminate other important quantum states, such as the vacuum and single photon superposition states.

We implement a displacement photon counter, characterize it using coherent state quantum detector tomography, and indirectly obtain the expected discrimination error probability by analyzing the tomographically reconstructed POVMs \cite{Lundeen09, Brida, Lita, Natarajan, Akhlaghi,Renema,Zhang12,catprojection}.
The characterization could also be done by probing directly with the $\ket{\pm}$ states, but preparing these with high quality is still technically challenging \cite{Imoto,Lombardi,Babichev}.
Reference \cite{Zhang12} first demonstrated the tomography of the displacement photon counter (referred to as weak homodyne detector), where a large detector Hilbert space is considered. 
Furthermore, the displacement photon counter was tomographically reconstructed in a two-dimensional space spanned by the superposition of the coherent state bases $|C_\pm\rangle \propto 
(|\alpha\rangle \pm |{-}\alpha\rangle)$ \cite{catprojection}.
Though our detection strategy is composed of the same resources, we restrict our focus to the two-dimensional space spanned by the vacuum and the single photon bases and the measurement is optimized in this space for minimization of the discrimination error for the superposition states.

\section{Displacement photon counter}\label{Sect:2}

\begin{figure}[b]
\centering
{
\includegraphics[width=0.9\linewidth]
{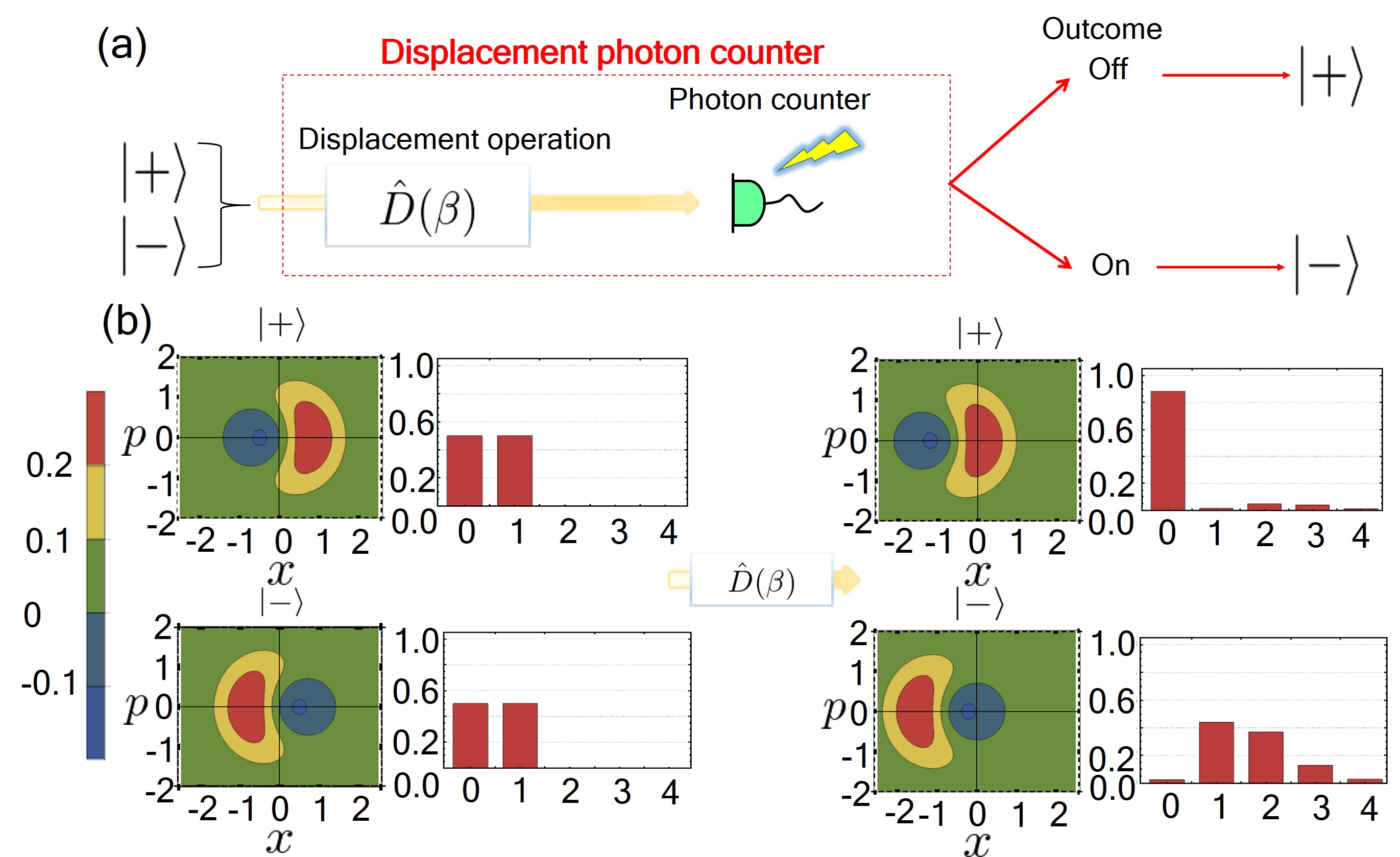}
}
\caption{(a) Schematic of the displacement photon counter.
(b) Phase space representation and photon number distribution of the superposition states. 
\label{kennedy}
}
\end{figure}

A schematic of the displacement photon counter is shown in Fig.~\ref{kennedy}(a).
It is composed of a displacement operation followed by a photon counter.
The phase space displacement operation displaces $\ket{+}$ close to a vacuum state and $\ket{-}$ to a finite amplitude state as shown in Fig.~\ref{kennedy}(b) after which the two states can be approximately distinguished by a photon counter: If the detector clicks, we conclude that the measured state is $\ket{-}$ while the absence of a click indicates that the state $\ket{+}$ was measured.   
For an ideal photon counter, the absence of a click is described by the projection onto vacuum, $|0 \rangle \langle 0|$.
Therefore the measurement operators (POVM elements) of the displacement photon counter are given by
\begin{eqnarray}
\hat{\Pi}_+^{\mathrm{K}} & = & \hat{D}(\beta)^{\dagger} |0 \rangle \langle 0| \hat{D}(\beta ),
\nonumber
\\
\hat{\Pi}_-^{\mathrm{K}} & = & \hat{I} - \hat{\Pi}_+^{\mathrm{K}},
\label{eq1}
\end{eqnarray}
with $\hat{D}(\beta)=\exp(\beta \hat{a}^{\dagger} - \beta^{\ast} \hat{a})$ being the displacement operation with amplitude $\beta$.
As the receiver is not optimal, errors can occur: An incoming $\ket{-}$ state can be mistaken for $\ket{+}$ and vice versa.
The average discrimination error is 
\begin{eqnarray}
P_e^K& = &\frac{\bra{-}\hat{\Pi}_+^{\mathrm{K}} \ket{-}+\bra{+}\hat{\Pi}_-^{\mathrm{K}} \ket{+}}{2}
\nonumber
\\
& = &
\frac{1}{2}+\mathrm{Re}[\beta]\mathrm{e^{-\abs{\beta}^2}},
\label{eq2}
\end{eqnarray}
where {\it a priori} probabilities are assumed to be equal. 
The minimum average error probability provided by the displacement photon counter is 
$P_e^K\approx 0.0711$ with the displacement amplitude $\beta=-1/\sqrt{2}$.
This optimal displacement value indicates that the displacement operation should be carefully adjusted so as to minimize the error probability rather than displacing $\ket{+}$ as close to the vacuum state as possible, which occurs for $\beta=-(\sqrt{5}-1)/2$.

It was shown that for BPSK signals, the minimum error probability under general Gaussian operations and adaptive feedback control with homodyne detection can be achieved by a simple homodyne 
detection \cite{Takeoka2008}.  
We want to benchmark our detector against an optimal Gaussian detector for discriminating $\ket{\pm}$. While the optimal Gaussian detector for discriminating BPSK coherent states is known \cite{Takeoka2008}, the optimal Gaussian detector for discriminating $\ket{\pm}$ and thus the minimum Gaussian error rate is not known to the best of our knowledge. 
In the following, we therefore consider a detection scheme consisting of a general Gaussian unitary transformation followed by a static homodyne detection and deduce the minimum error rate which will be serving as the benchmark for our detector.
Any single mode Gaussian operation $\hat{U}_G$ can be decomposed as $\hat{U}_G=\hat{D}(\alpha)\hat{R}(\phi)\hat{S}(r)\hat{R}(\theta)$, where $\hat{R}(\cdot)$ and $\hat{S}(r)$ represent a phase rotation operator and a single mode squeezing operator, respectively \cite{gaussian_review}. Therefore, the POVM of a general Gaussian operation followed by a static homodyne detection with the measurement outcome $x$ is described as 
$\{\hat{\Pi}_x^G=\hat{U}_G^{\dagger} |x \rangle \langle x| \hat{U}_G\}_{x \in \mathbb{R}}$.
The matrix representation of the POVM in the two-dimensional space of interest is
\begin{equation}
\label{eq:gauss_twoD}
\left[
\begin{array}{cc}
\bra{0}\hat{\Pi}_x^G \ket{0} & \bra{0}\hat{\Pi}_x^G \ket{1} \\
\bra{1}\hat{\Pi}_x^G \ket{0} & 
\bra{1}\hat{\Pi}_x^G \ket{1}
\end{array}
\right] =
\mathcal{N} \left[
\begin{array}{cc}
1 & \epsilon\tilde{x} \\
\epsilon^{\ast}\tilde{x} & \abs{\epsilon}^2\tilde{x}^2
\end{array}
\right],
\end{equation}
where
\begin{eqnarray}
\mathcal{N}&=&\frac{\mathrm{sech}r}{\sqrt{\pi}\abs{1-\xi}^2}
\exp{[-\frac{\mathrm{sech}^2r}{\sqrt{\pi}\abs{1-\xi}^2}\tilde{x}^2]},
\\
\epsilon&=&\frac{\mathrm{e}^{i (\phi+\theta)}\sqrt{2}\mathrm{sech}r}{1-\xi},
\\
\xi&=&\mathrm{e}^{2i\phi}\tanh{r},
\\
\tilde{x}&=&x-\sqrt{2}\mathrm{Re}[\alpha].
\end{eqnarray}
By comparing {\it a posteriori} probability distributions $\tr{[\bra{\pm}\hat{\Pi}_x^G\ket{\pm}]}$,
we find that the homodyne outcomes should be distributed to a binary decision according to
$\{\hat{\Pi}_+^{\mathrm{G}}=\int_{0}^\infty \hat{\Pi}_x^G d\tilde{x}, \hat{\Pi}_-^{\mathrm{G}}=\hat{I}-\hat{\Pi}_+^{\mathrm{G}} \}$
for $\mathrm{Re}[\epsilon] \geq 0$ and oppositely
for $\mathrm{Re}[\epsilon] < 0$.
The average error probability, defined equivalently to Eq. (\ref{eq2}), is then
\beq 
P_e^G=
\frac{1}{2}-\frac{1}{\sqrt{2\pi}} \frac{\abs{\cos{(\theta+\phi)}-\cos{(\theta-\phi)}\tanh{r} }} {\sqrt{1-2\cos{2\phi}\tanh{r}+\tanh^2{r}}}.
\label{eq3}
\eeq
The second term is maximized when $\{\theta,\phi\}\rightarrow\{0,0 \}$ and for any $r$, i.e., the minimum error probability for the general Gaussian unitary operations followed by the static homodyne detection is achievable by a simple homodyne detection and its error probability is $P_e^G = \frac{1}{2} - \frac{1}{\sqrt{2\pi}} \approx 0.101$.


\section{Experiment}\label{Sect:3}
\begin{figure}[b]
\centering
{
\includegraphics[width=0.95\linewidth]
{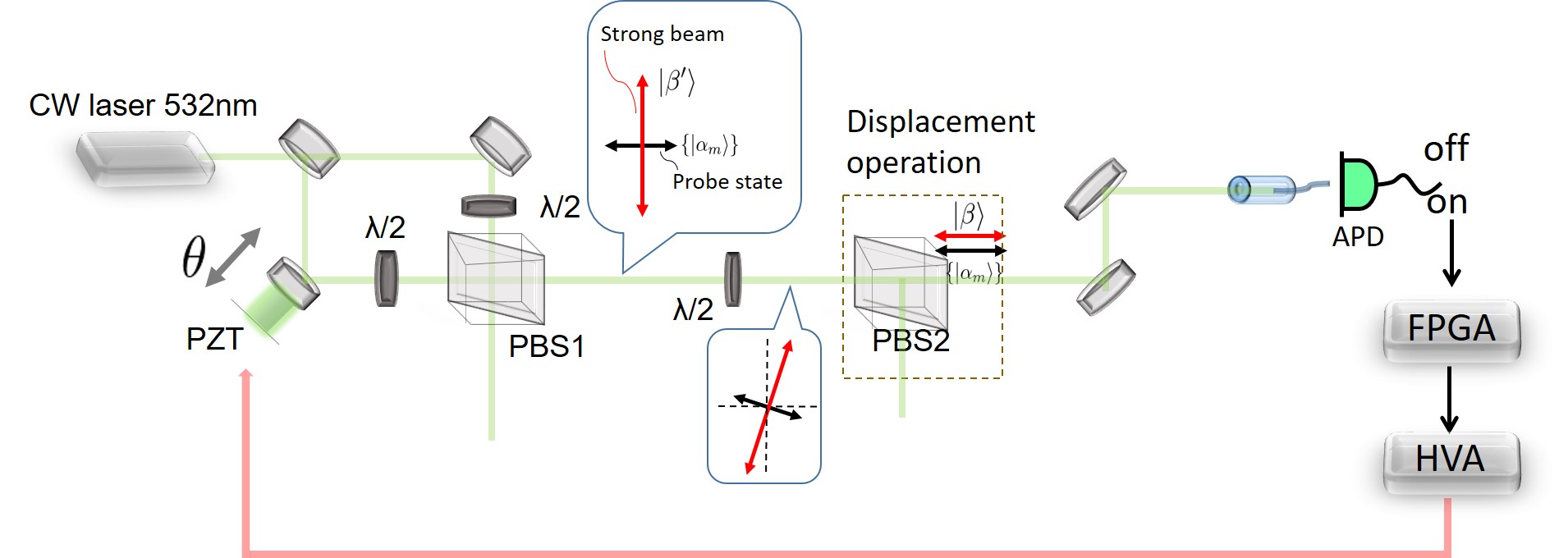}
}
\caption{
Experimental setup.
PBS: 
polarization beam splitter,
APD:
avalanche photo diode,
PZT:
piezo transducer,
FPGA:
field programmable gate array,
HVA: 
high voltage amplifier.
}
\label{Experiment_setup}
\end{figure}
In our experiment, we implement a displacement photon counter and characterize it using quantum detector tomography where a large number of probes, selected from a small set of coherent states that cover the Hilbert space of interest, are injected into the receiver. From the distribution of detection outcomes, the POVM of the measurement is reconstructed following a maximum likelihood (ML) procedure \cite{Fiurasek,Paris_book}. 
The expected error rate for discrimination of the superposition states is then estimated by
\begin{equation}
P_e = \frac{\bra{-}\hat{\Pi}_+^{\mathrm{ex}} \ket{-}+\bra{+}\hat{\Pi}_-^{\mathrm{ex}} \ket{+}}{2},
\label{eq4}
\end{equation}
where $\hat{\Pi}_{\pm}^{\mathrm{ex}}$ represent the experimentally reconstructed POVM elements.
The ML is dedicated to find the POVMs that maximize the log-likelihood functional defined as
\begin{equation}
{\mathcal L}[\{\hat{\Pi}_l \}]=\sum_{l=1}^{L} \sum_{m=1}^{M} f_{ml} \ln{\tr{[\hat{\rho}_m \hat{\Pi}_l]}},
\label{eq10}
\end{equation}
under the constraints for the POVMs, $\{ \hat{\Pi}_l\geq 0, \sum_{l=1}^{L} \hat{\Pi}_l =\hat{I} \}$.
$L, M, f_{ml}$ indicate the number of POVMs to be estimated, the number of probe states and
the experimentally obtained frequency of the outcome $l$ when probing with the state $\hat{\rho}_m$.
The most likely POVMs maximizing Eq.~(\ref{eq10}) for the acquired data can be attained by recursively applying the following transformation,
\begin{equation}
\hat{\Pi}_l= \hat{\lambda}^{-1} \hat{R}_l \hat{\Pi}_l \hat{R}_l \hat{\lambda}^{-1},
\end{equation}
where
\begin{eqnarray}
\hat{R}_l&=& \sum_{m=1}^{M} \frac{f_{ml}}{\tr{[\hat{\rho}_m \hat{\Pi}_l]}}\hat{\rho}_m, 
\\
\hat{\lambda}&=& (\sum_{l=1}^L \hat{R}_l \hat{\Pi}_l \hat{R}_l)^{1/2}.
\end{eqnarray}
For the tomography in our experiment,
we choose the amplitudes of the probe states such that they can be approximately represented in the four-dimensional space spanned by photon numbers 0--3 and fully cover the two-dimensional space of interest.
Our measurement provides binary outcomes $\{\hat{\Pi}_{+}, \hat{\Pi}_{-}\}$ ($L=2$) and
we prepare 16 different probe coherent states ($M=16$) with 4 different mean photon numbers and phases $\{\pi/4, 3\pi/4, 5\pi/4, 7\pi/4 \}$.

Figure \ref{Experiment_setup} shows our experimental setup.
We use a continuous wave laser at 532 nm and 
the temporal mode of the coherent probe states is defined as a 1 $\mathrm{\mu}$s segment of the continuous wave beam. 
The displacement operation of the receiver can be physically implemented by combining the optical state with a strong beam on a highly transmitting beam-splitter.
For this purpose, the beam is split in two paths: 
In one path, the coherent probe states are prepared. 
Their phase and amplitude are controlled by a piezo-mounted mirror and a half-wave plate plus the PBS1 polarizing beam-splitter, respectively. 
The other path is used as the strong displacement beam. 
These two beams are spatially combined with orthogonal polarizations at PBS1 and are made to interfere at the subsequent PBS2. 
To effectively obtain a highly transmitting beam-splitter operation, the polarization of the two beams is rotated just slightly before PBS2.
The displaced probe state is guided into a single mode fiber and detected by an avalanche photo diode (APD).
We develop a field programmable gate array (FPGA) to count the electric signals from the APD.
The individual clicks (or non-clicks) are the outcomes of the state measurements, while the overall count rate is used for feedback control of the relative phase between probe and displacement through an applied voltage on the piezo mirror.
The optical losses due to fiber coupling and non-unit detection efficiency of the APD with a total effective efficiency of $\eta$ cause a degradation of the performance of the displacement photon counter.
However, the optical losses in our experiment can be compensated by rescaling the amplitude of the coherent state $|\sqrt{\eta}\alpha\rangle\rightarrow \ket{\alpha}$
since the coherent state remains a pure state and the phase is invariant under linear losses.
We are thus characterizing the performance of an idealized, perfectly efficient receiver since we are interested in the ultimately achievable discrimination ability.
In practice, this loss compensation happens automatically due to our calibration of the probe state and displacement amplitudes through the observed photon count rates.

\begin{figure}[t]
\centering
{
\includegraphics[width=\linewidth]
{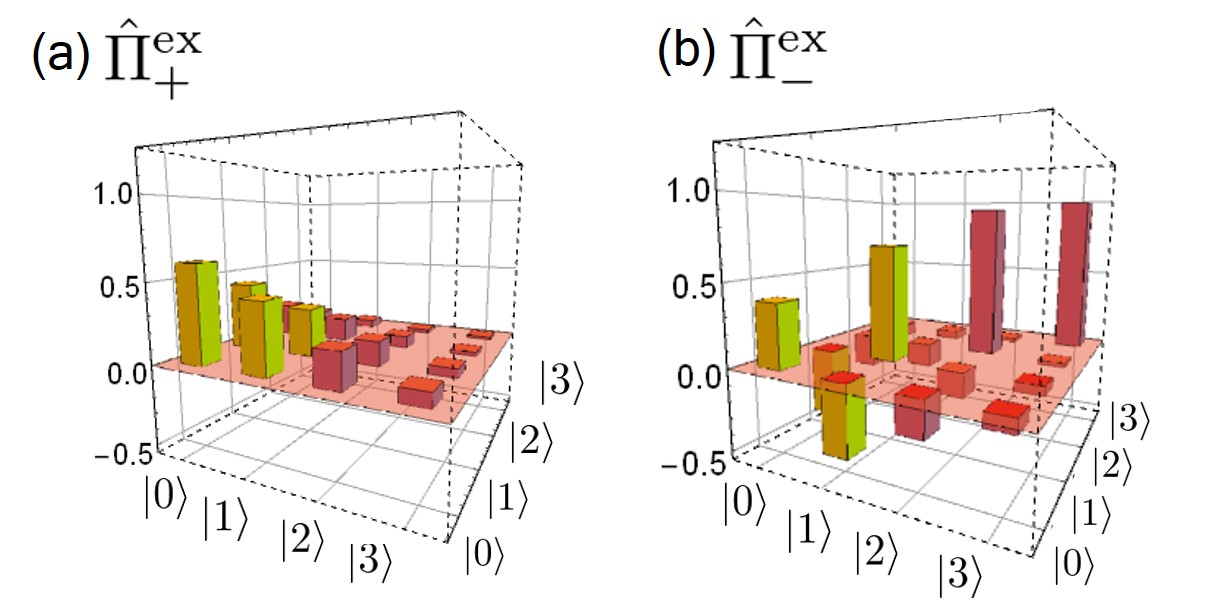}
}
\caption{
Real part of the reconstructed POVMs of the displacement photon counter, 
(a) $\hat{\Pi}_+^{\mathrm{ex}}$,
(b) $\hat{\Pi}_-^{\mathrm{ex}}$. 
The amplitude of the displacement operation is set to $\beta=-0.70$.
The yellow bars indicate the truncated two-dimensional space.
}
\label{Kennedy_POVM}
\end{figure}


The real parts of the reconstructed POVM elements are depicted in Fig.~\ref{Kennedy_POVM},
where the amplitude of the displacement operation is set to $\beta=-0.70\pm0.02$.
$5\times 10^4$ experimental data are acquired for each probe state.
The imaginary parts of the off diagonal elements in the reconstructed POVM that originate from a misaligned phase of the displacement operation are negligibly small.
The original POVM is reconstructed in the four-dimensional space in the photon number basis and truncated to the two-dimensional space in order to evaluate the error probability. 
The fidelity between the POVM elements of the displacement photon counter given in Eq.~(\ref{eq1}) and the experimentally reconstructed ones with the two-dimensional truncation can be defined as \cite{Zhang12},  
\begin{equation}
\mathcal{F}_{\pm}=(\tr{[(\sqrt{\hat{\Pi}^{\rm{K}}_{\pm} }\hat{\Pi}^{\rm{ex}}_{\pm} \sqrt{\hat{\Pi}^{\rm{K}}_{\pm} })^{1/2}]})^2/(\tr{[\hat{\Pi}^{\rm{ex}}_{\pm} ]}\tr{[\hat{\Pi}^{\rm{K}}_{\pm} ]}).
 \label{eq8_4}
\end{equation}
Both fidelities surpass 99.5\%.

\begin{figure}[t]
\centering
{
\includegraphics[width=0.95\linewidth]
{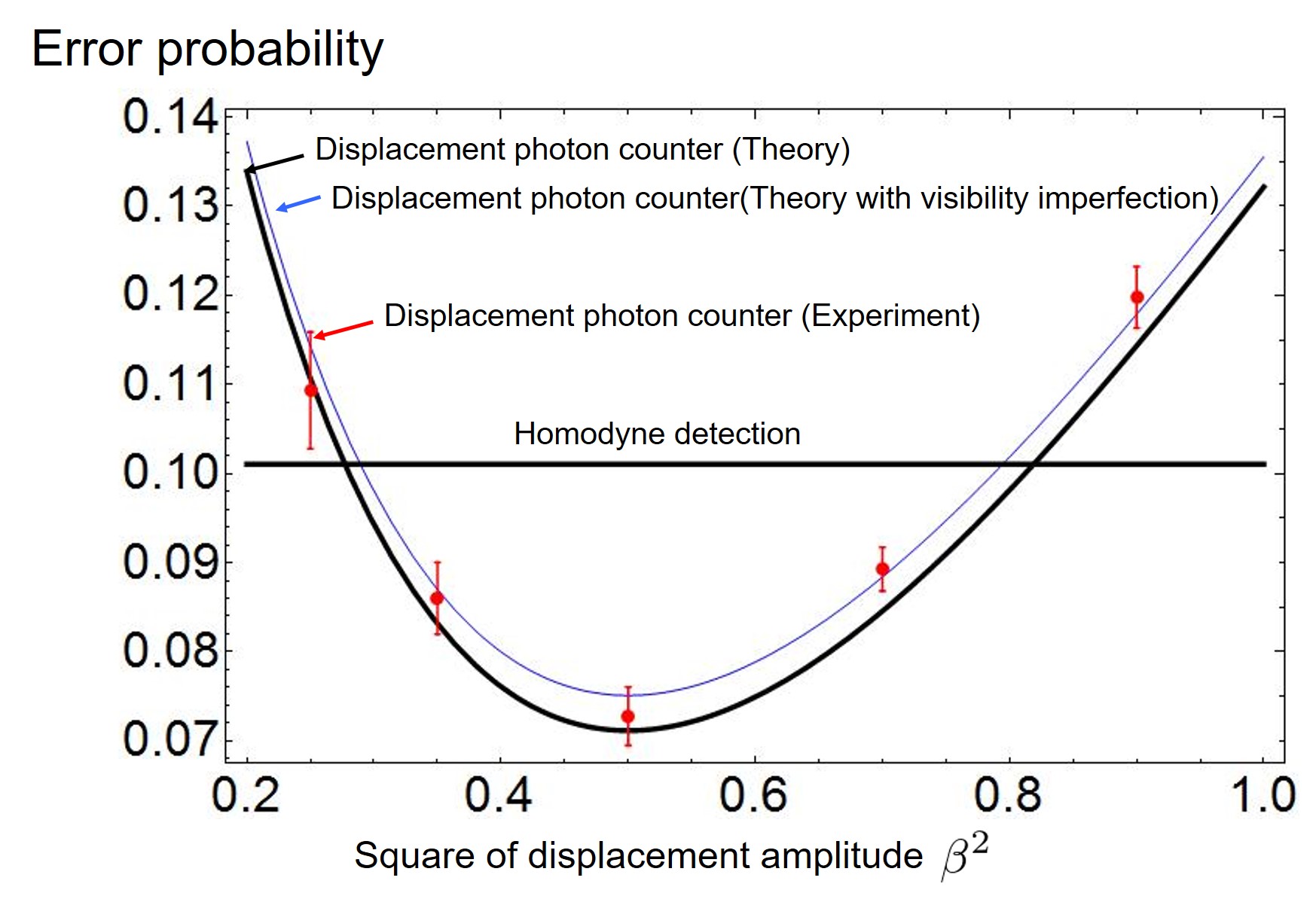}
}
\caption{
Achievable error probability calculated from the experimentally reconstructed POVMs (red points), the theoretical prediction of the error probability for the displacement photon counter (black curve), the theoretical prediction including the observed non-perfect visibility (thin blue curve), and the homodyne detection (horizontal black line).
}
\label{Error_experiment}
\end{figure}

In Fig.~\ref{Error_experiment}, we evaluate the achievable error probability obtained from the experimentally realized measurement.
We examine 5 displacement amplitude conditions.
The average values and error bars are obtained from 5 independent procedures.
The experimental result agrees well with the theoretical prediction (black curve) and beats the performance of the homodyne detection (black horizontal line) for an optimized amplitude of the displacement operation.
The agreement is even better if the model takes into account the visibility imperfection of the displacement operation and the dark counts of the APD (blue curve), which are measured to be $99.1 \%$ and $310$Hz respectively.

\section{Conclusion}\label{Sect:4}
We applied the displacement photon counter, which is known as a near optimal measurement strategy for the BPSK coherent states discrimination, to the discrimination of equally weighted superpositions of the vacuum and the single photon states -- the $\hat{\sigma}_x$ eigenstates in the single-rail qubit encoding.
We experimentally realized the displacement photon counter and characterized our measurement by quantum detector tomography with coherent probe states.
The achievable error probability was indirectly evaluated with the reconstructed POVMs.
Our proof-of-principle experiment showed that the displacement photon counter provides a better error probability than the homodyne detection by optimizing the displacement amplitude. 

An interesting remaining issue is a physical implementation of the projection measurement that can perfectly discriminate the superposition states.
Indeed, as shown in Ref.~\cite{TakeokaSasakiLutkenhaus2006_PRL_BinaryProjMmt}, an error-free discrimination of the superposition states is achievable, not only for equally weighted but also for arbitrary orthogonal superpositions, by applying an infinitely fast electrical feedback operation to the displacement photon counter, i.e., the Dolinar receiver.
Such a two-dimensional projection measurement is an essential tool for quantum information processing with single-rail optical qubits where an arbitrary two-dimensional projector in the qubit Hilbert space is required.
Although there still exists a gap between the displacement photon counter and the ideal projection measurement, 
our work shows the potential of well-established techniques for the single-rail qubits discrimination.
\begin{acknowledgements}
This project was supported by the VILLUM FOUNDATION Young Investigator Programme and Grant-in-Aid for JSPS Research Fellow.
\end{acknowledgements}


\end{document}